\documentclass[12pt]{iopart}
\usepackage{iopams}  
\usepackage{cite}  
\usepackage[dvips]{graphics}
\usepackage{epsfig}

\begin{document}

\title[]{Particle-in-cell simulations of high energy electron production by intense laser pulses in underdense plasmas}

\author{{Susumu Kato}\dag
, Eisuke Miura\dag, Mitsumori Tanimoto\ddag, Masahiro Adachi\S, Kazuyoshi Koyama\dag}

\address{\dag\ National Institute of Advanced Industrial Science and Technology (AIST), Tsukuba, Ibaraki 305-8568, Japan}

\address{\ddag\ Department of Electrical Engineering, Meisei University, Hino, Tokyo 191-8506, Japan}

\address{\S\ Graduate school of Advanced Science of Matter, Hiroshima University, Higashi-Hiroshima, Hiroshima 739-8530, Japan}

\begin{abstract}
The propagation of intense laser pulses and the generation of high energy electrons from the underdense plasmas are investigated using two dimensional particle-in-cell simulations.
When the ratio of the laser power and a critical power of relativistic self-focusing is the optimal value, it propagates stably and electrons have maximum energies.
\end{abstract}





\section{Introduction}
The interaction of an intense ultrashort laser pulse with underdense plasmas has attracted much interest for a compact accelerator. Using intense laser systems, of which peak powers exceed 10 TW, electrons with energies greater than 100 MeV have been observed in various low density range, of which electron densities are from $6\times10^{16} \mathrm{cm}^{-3}$ to $2.5\times10^{19} \mathrm{cm}^{-3}$ \cite{Najmudin2003,kitagawa2004}.
On the other hand, using terawatt class Ti:sapphire laser systems, electrons with energies greater than several mega-electron-volts have been observed from moderately underdense plasmas, of which densities are up to near the quarter-critical density\cite{gahn1999,miura2003}. At the moderately underdense plasmas, the electron energies exceed the maximum energies determined by dephasing length. It is considered recently that the acceleration occurs by the direct laser acceleration\cite{tanimoto2003} that includes stochastic or chaotic processes.

In this paper, we study the propagation of the laser pulses and the generation of high energy electron in the underdense plasmas using two dimensional particle-in-cell (2D PIC) simulations.
The laser power $P_{\mathrm{L}}$ beyond the critical power $P_{\mathrm{cr}}$ is necessary because self-focusing is important in a long-distance propagation\cite{schmidt1985}. Here, $P_{\mathrm{cr}}\simeq17\left(n_\mathrm{c}/n_\mathrm{e}\right)\mathrm{GW}$, $n_\mathrm{e}$ and $n_\mathrm{c}$ are the electron density and critical density, respectively.
We assume a terawatt class Ti:sapphire laser system as a compact laser system in the simulations parameter, because the plasma electron densities $n_e=(1\sim20)\times10^{19}\textrm{cm}^{-3}$.

\section{2D PIC Simulation Results}

We use the 2D PIC simulation with immobile ions. The peak irradiated intensity, pulse length, and spot size are $5\times10^{18}$ $\textrm{W/cm}^2$, $50$ fs, and 3.6 $\mu$m, respectively. $P_{\mathrm{L}} = 2$ TW, namely the energy is $100$ mJ, when cylindrical symmetry is assumed, although we use two dimensional cartesian coordinates. The Rayleigh length $L_\mathrm{R}=50 \mu$m.
The plasma electron densities $n_e=(1\sim20)\times10^{19}\textrm{cm}^{-3}$, which correspond to $n_\mathrm{e}/n_\mathrm{c}=0.057\sim0.11$, where $n_\mathrm{c}=1.7\times10^{21}\textrm{cm}^{-3}$ for the wavelength $\lambda_0=0.8\mu\textrm{m}$.
These parameters of the simulations are almost the same as the experiments of compact laser system\cite{miura2003}. 
The laser power $P_{\mathrm{L}} = 2$ TW exceeds the critical powers $P_{\mathrm{cr}}$ of the relativistic self-focusing for $n_e\geqslant1\times10^{19}\textrm{cm}^{-3}$. 

Figures 1(a)-(e) show the intensities of laser pulses after propagating $2.5L_\mathrm{R}$ for electron densities $n_e=20, 10, 5, 2$, and $1\times10^{19}\textrm{cm}^{-3}$, respectively. For $n_\mathrm{e}=1$ and $2\times10^{19}\textrm{cm}^{-3}$, namely, $P_{\mathrm{L}}/P_{\mathrm{cr}} \simeq 1$, the pulses stably propagate without modulation. Electrons with energies greater than 2 MeV are hardly observed. For $n_\mathrm{e}=5\times10^{19}$$\textrm{cm}^{-3}$, the back of pulse is modulated. Electrons get energies greater than 20 MeV, as shown later. For $n_\mathrm{e}=1\times10^{20}$$\textrm{cm}^{-3}$, a pulse separates into the bunches of which size about the plasma wavelength. A pulse breaks up and is not propagate stably any longer, for $n_\mathrm{e}=2\times10^{20}$$\textrm{cm}^{-3}$, i.e. $P_{\mathrm{L}}/P_{\mathrm{cr}} \geqslant 10$. 
 
The electron energy spectra for electron densities $n_\mathrm{e}=20, 10$, and $5\times10^{19}\textrm{cm}^{-3}$ are shown in Figs. 2(a)-(c), respectively. The maximum energy is greater than $20$ MeV for $n_\mathrm{e}=5\times10^{19}\textrm{cm}^{-3}$. 
Before the pulses propagate about one and two Rayleigh length, the maximum electron energies have been saturated at 20 MeV and 10 MeV for $n_\mathrm{e}=1$ and $2\times10^{20}\textrm{cm}^{-3}$, respectively.

\section{Concluding Remarks}
We study the propagation of the intense laser pulses and the generation of high energy electrons from the moderately underdense plasmas using 2D PIC simulations. For $P_{\mathrm{L}}/P_{\mathrm{cr}} \simeq 3$, the laser pulse of which power and pulse length are $2$ TW and $50$ fs stably propagates with modulation. As a result, the high energy electrons with energies greater than 20 MeV are observed and their energies have not been saturated, namely, electrons can gain higher energies propagating with the intense laser pulse through long size plasmas. For $P_{\mathrm{L}}/P_{\mathrm{cr}} \leqslant 2$, although the pulses stably propagate, no high energy electron is observed. On the other hand, for $P_{\mathrm{L}}/P_{\mathrm{cr}} \geqslant 5$, high energy electrons with energies up to 20 MeV are observed, although pulses does not propagate stably.

The simulation results of the dependence to the plasma density of the maximum electron energy explain the latest experiment well qualitatively\cite{miura2003}. In the simulations, the maximum propagation distance is $3L_\mathrm{R}$ is limited by the performance of the computer and simulation code. Since the pulse has propagated  sufficiently stably to $2.5L_\mathrm{R}$ for the plasma densities less than $5\times10^{19}$$\textrm{cm}^{-3}$, simulations with a longer propagation distance is required. 

\ack {A part of this study was financially supported by the Budget for Nuclear Research of the Ministry of Education, Culture, Sports, Science and Technology (MEXT), based on the screening and counseling by the Atomic Energy Commission, and by the Advanced Compact Accelerator Development Program of the MEXT.}

\newpage
\begin{figure}[htbp]
\begin{center}
  \includegraphics[width=170mm]{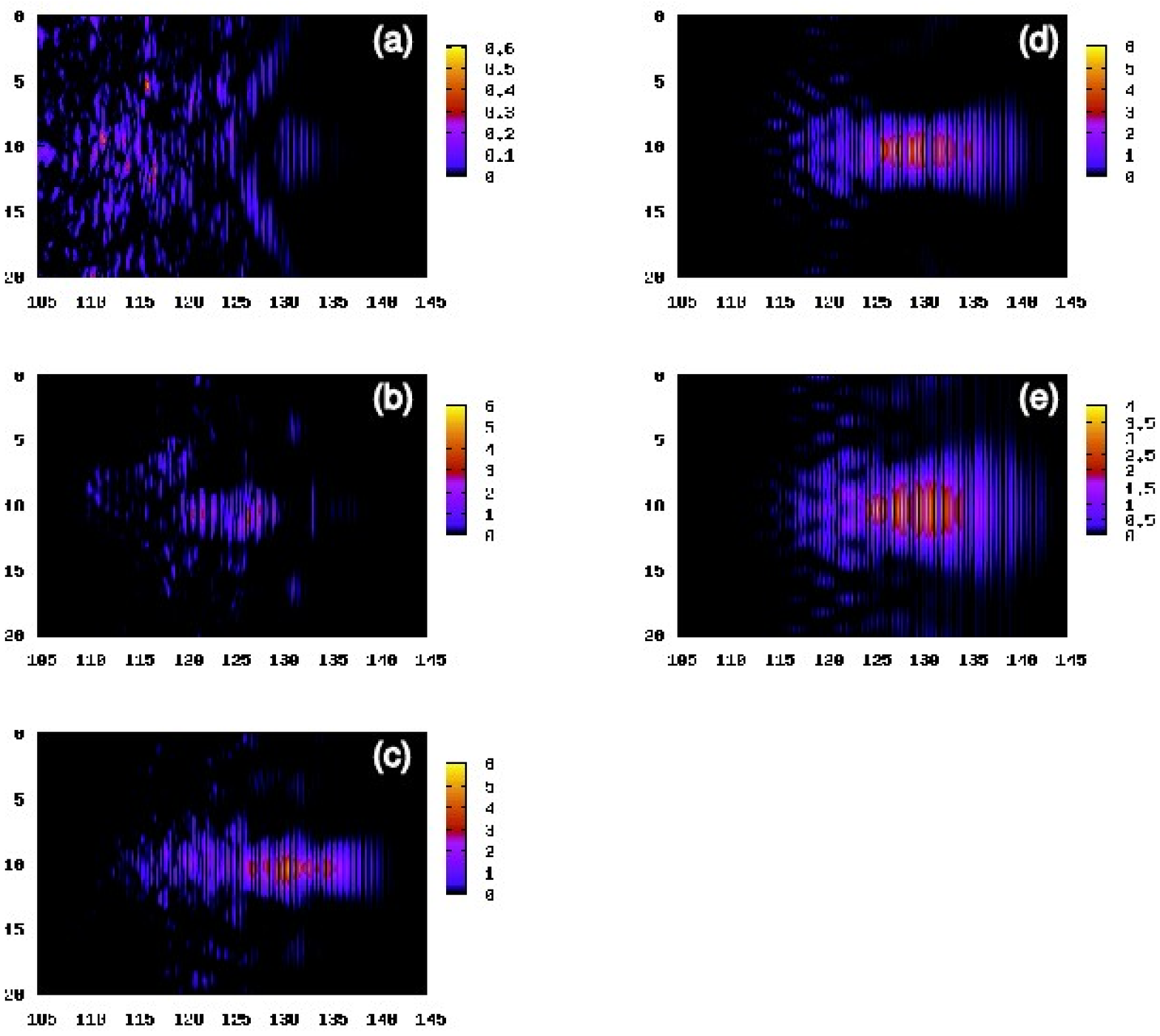}
\end{center}
\end{figure}
\Figure{\label{label1} The intensities of laser pulses after propagating $2.5L_\mathrm{R}$ for electron densities $n_e=$ (a) 20, (b) 10, (c) 5, (d) 2, and (e) $1\times10^{19}\textrm{cm}^{-3}$, respectively.}

\newpage
\begin{figure}[htbp]
\begin{center}
\includegraphics[width=80mm]{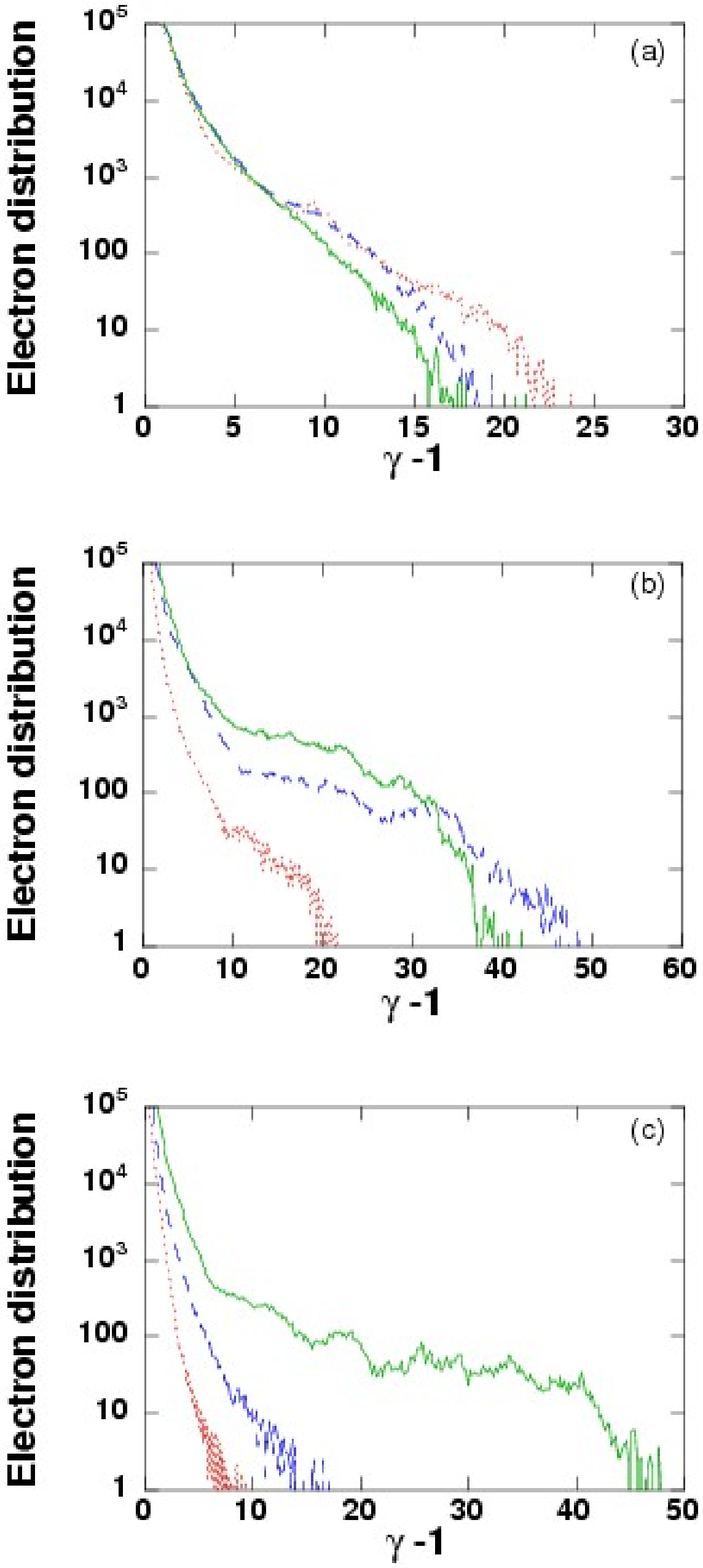}
\end{center}
\end{figure}
\Figure{\label{label2} Electron energy spectra for electron densities $n_\mathrm{e}=$ (a) $20$, (b) 10, and (c) $5\times10^{19}\textrm{cm}^{-3}$, respectively. The dotted, dashed, and solid lines are after propagating 1, 2, and 3$L_\mathrm{R}$, respectively }

\end{document}